\documentclass[prl,aps,twocolumn,superscriptaddress]{revtex4}

\usepackage{bm}
\usepackage{graphicx}
\usepackage{amsbsy}
\usepackage{amsmath}
\usepackage{amsfonts}
\usepackage{amsthm}
\usepackage{listings}

\begin{document}

\theoremstyle{plain}
\newtheorem{theorem}{Theorem}
\newtheorem{lemma}[theorem]{Lemma}
\newtheorem{corollary}[theorem]{Corollary}
\newtheorem{conjecture}[theorem]{Conjecture}
\newtheorem{proposition}[theorem]{Proposition}

\theoremstyle{definition}
\newtheorem{definition}{Definition}

\theoremstyle{remark}
\newtheorem*{remark}{Remark}
\newtheorem{example}{Example}

\def\be{\begin{equation}}
\def\ee{\end{equation}}
\def\ba{\begin{align}}
\def\ea{\end{align}}

\newcommand{\mE}{\mathcal{E}}
\newcommand{\mU}{\mathcal{U}}
\newcommand{\mA}{\mathcal{A}}
\newcommand{\mF}{\mathcal{F}}

\newcommand{\fm}{\mathcal{F}_{\bf{m}}}
\newcommand{\am}{\mathcal{A}^{\textbf{m}}}
\newcommand{\dm}{\mathcal{D}(\mathrm{H}_{{\bf m}})}

\newcommand{\mbN}{\mathbb{N}}

\title{A Reversible Framework for Resource Theories}   
\author{Fernando G.S.L. Brand\~ao}\email{fgslbrandao@gmail.com}
\affiliation{Quantum Architectures and Computation Group, Microsoft Research, Redmond, WA}
\affiliation{Department of Computer Science, University College London}
\author{Gilad Gour}\email{gour@ucalgary.ca}
\affiliation{
Department of Mathematics and Statistics,
University of Calgary, 2500 University Drive NW,
Calgary, Alberta, Canada T2N 1N4}

\date{\today}


\begin{abstract}
In recent years it was recognized that properties of physical systems such as entanglement, athermality, and asymmetry, can be viewed as resources for important tasks in quantum information, thermodynamics, and other areas of physics. This recognition followed by the development of specific quantum resource theories (QRTs), such as entanglement theory, determining how quantum states that cannot be prepared under certain restrictions may be manipulated and used to circumvent the restrictions. Here we discuss the general structure of QRTs, and show that under a few assumptions (such as convexity of the set of free states), a QRT is asymptotically reversible if its set of allowed operations is \emph{maximal}; that is, if the allowed operations are the set of all operations that do not generate (asymptotically) a resource.
In this case, the asymptotic conversion rate is given in terms of the regularized relative entropy of a resource which is the unique measure/quantifier of the resource in the asymptotic limit of many copies of the state. This measure also equals the smoothed version of the logarithmic robustness of the resource.
\end{abstract}

\maketitle


Classical and quantum information theories can be viewed as examples of theories of interconversions among different resources~\cite{DW04}.
These resources are classified as being quantum or classical, dynamic or static, noisy or noiseless, 
and therefore enable a plethora of quantum information processing tasks~\cite{NC11,W13}. For example, quantum teleportation can be viewed as a resource interconversion task in which
one entangled bit (a quantum static noiseless resource) 
is transformed by local operations and classical communication (LOCC) into a single use of a quantum channel (a quantum dynamic noiseless resource)~\cite{PV07}. 
Just as the restriction of LOCC leads to the theory of entanglement~\cite{HHH09}, 
in general, every restriction on quantum operations defines a resource theory, determining how quantum states that cannot be prepared under the restriction may be manipulated and used to circumvent the restriction. 

The scope of quantum resource theories (QRTs) goes far beyond quantum information science~\cite{HOO2013}. In recent years
a lot of work has been done formulating QRTs in different areas of physics, such as the resource theory of athermality in quantum thermodynamics~\cite{BHO13,BHN13,HO13,FDO12,GMN14,NG14}, the resource theory of asymmetry~\cite{Collection,MS14}
(which led to generalizations of important theorems in physics such as Noether's theorem~\cite{MS14}), the resource theory of non-Gaussianity in quantum optics~\cite{BL05,BJS03}, the resource theory of stabilizer computation in quantum computing~\cite{VMG14}, non-contextuality in the foundations of quantum physics~\cite{GHH14}, and more recently it was suggested that non-Markovian evolution can be formulated as a resource theory~\cite{RHP14}. In addition, tools and ideas from quantum resource theories have been applied in many-body physics (see e.g.~\cite{BH13} and references therein), and even for a universal formulation of the uncertainty principle~\cite{FGG13}. Furthermore, very recently an abstract formulation using concepts from category theory has been proposed, unifying all resource theories into a single framework~\cite{CFS14}.
 
Despite this large body of work, so far there are no known theorems that can be applied to a large class of QRTs. In this Letter we prove one such theorem, establishing a criterion of when a resource theory is asymptotically reversible. In particular, we show that under a few physically motivated assumptions, a resource theory is asymptotically reversible if its set of allowed operations is \emph{maximal}; that is, if the allowed operations are the set of all operations that do not generate (asymptotically) a resource. Our approach is a generalization of the results of~\cite{BP08} from entanglement
to general resource theories satisfying a few basic properties. Our main
innovation is to show that the arguments of~\cite{BP08} can be extended to resource
theories where there is no notion of a maximally valuable resource, as in
the case of entanglement theory. Thus, our work also simplifies parts of the
proof in~\cite{BP08}.

QRTs have a general structure; they all consists of three main ingredients: 
(1) the resources (like entanglement), (2) the non-resources or free states (like separable non-entangled states in entanglement theory), 
and (3) the restricted set of free (or allowed) operations (like LOCC in entanglement theory). This structure give rise to two extreme limits corresponding to trivial resource theories. 
In the first one, the restriction is very loose and almost nothing is a resource since almost every operation is allowed. 
The other extreme limit is when the restriction is very strong and almost every quantum state is a resource since it can not be prepared under the set of allowed operations.
The most interesting resource theories are those for which the restrictions on the allowed operations are somewhere in the middle of these two extremes.
An important point between these two limits is when the restriction is strong enough so that the theory is not trivial and yet loose enough so that the resource theory is asymptotically reversible.

The three constituents of a resource theory -- namely the free states, the allowed/restricted operations, and the resources -- are not independent of each other.
For example, the restricted set of operations must be such that it does not generate resources from free states (otherwise, it can not be called a resource theory). 
Therefore, any assumption being made on one of these ingredients effects the others. Below we give 5 physically motivated postulates on the set of free states 
that will be used to prove our main result. 

All systems considered here are finite dimensional, so that for every system, described by 
a state $\rho$, there exists integer 
$s\geq 2$ and ${\bf{m}}\equiv(m_1,...,m_s)$ (with $m_j$ positive integers) such that
$\rho\in\mathcal{D}(\mathrm{H}_{{\bf m}})$, where $\mathrm{H}_{{\bf m}}\equiv\mathbb{C}^{m_1}\otimes\mathbb{C}^{m_2}\otimes\cdots\otimes\mathbb{C}^{m_s}$
and $\mathcal{D}(\mathrm{H}_{{\bf m}})$ is the convex set of density matrices acting on $\mathrm{H}_{{\bf m}}$. We denote by $\mathcal{F}$ the set of all free states (in all possible finite dimensions), and by $\mathcal{F}_{\bf{m}}=\mathcal{F}\cap\mathcal{D}(\mathrm{H}_{{\bf m}})$ the free states in 
$\mathcal{D}(\mathrm{H}_{{\bf m}})$.
The free states are states that can be generated freely at no cost. Therefore, if a state $\sigma\in\mathcal{D}((\mathrm{H}_{{\bf m}}))$  
is free so is $\sigma\otimes\sigma$.
We conclude that if $\rho,\sigma\in\mathcal{F}$ then $\rho\otimes\sigma\in\mathcal{F}$.
We summarize this with the first postulate on $\mathcal{F}$:

\textbf{Postulate I}: {\it The set of free states $\mathcal{F}$ is closed under tensor products.}

The second postulate is the converse of the first postulate. That is, if $\sigma\in \mathcal{D}\left(\mathrm{H}_{{\bf m}}\otimes \mathrm{H}_{{\bf m'}}\right)$
represents a composite bipartite system, then discarding one of the subsystems can not generate a resource. 
We will only assume that it is possible to discard a subsystem at no cost if the subsystems are spatially separated. This amounts mathematically to the partial trace. Note however that for a \emph{single} system partial trace will not be allowed even if the Hilbert space of the single system is isomorphic to a tensor product of Hilbert spaces.

\textbf{Postulate II}: {\it The set of free states $\mathcal{F}$ is closed under the partial trace of spatially separated subsystems.}

Clearly, in any reasonable resource theory if $\rho$ and $\sigma$ are free states then both $\rho\otimes\sigma$ and $\sigma\otimes\rho$ should be free. Taking this one step further, we will assume that if a free state $\rho\in\mathcal{D}\left(\mathbb{C}^{m_1}\otimes\cdots\otimes\mathbb{C}^{m_s}\right)$ represents a composite system with $s$ spatially separated subsystems, then the permutation of the $s$ subsystems can not generate a resource.

\textbf{Postulate III}: {\it The set of free states $\mathcal{F}$ is closed under permutations of spatially separated subsystems.}

The next postulate concerns with continuity. If a sequence of \emph{free} states $\{\rho_n\}$ converges to a state $\rho$ (with respect to any of the $\ell^p$-norms; i.e. $\lim_{n\to\infty}\|\rho_n-\rho\|_p=0$) then the state $\rho$ must also be free. Otherwise, the resource theory will not be continuous. 

\textbf{Postulate IV}: {\it Each $\mathcal{F}_{\bf{m}}$ is a closed set.}

The next postulate concerns with convexity.  
Suppose $\rho$ and $\sigma$ are two free states both acting on the same Hilbert space, and suppose one (say Alice) flips an unbiased coin (assuming such a coin is by itself not a resource and available to Alice). If Alice gets a head then she prepares $\rho$ and if she gets a tail then she prepares $\sigma$. Here we assume that if Alice forgets whether she got a head or a tail that alone can not generate a resource. That is, ${1\over 2}\rho+{1\over 2}\sigma$ should also be a free state. In the same way, since both $\rho$ and ${1\over 2}\rho+{1\over 2}\sigma$ are free states so is $\frac{3}{4}\rho+\frac{1}{4}\sigma$. Continuing in this way, we get that ${k\over 2^n}\rho+\left(1-{k\over 2^n}\right)\sigma$ is a free state for all $n\in\mathbb{N}$ and $k=0,1,2,...,2^n$. Since the set $\left\{k\over 2^n\right\}$ is dense in $[0,1]$ the previous postulate implies that for any $t\in[0,1]$ the state $t\rho+(1-t)\sigma$ is free. Note that we arrived at this conclusion assuming one has access to randomness, i.e. the unbiased coin (also biased coins will do the job), as well as free classical communication in distributed settings. Clearly, in some QRTs these assumptions don't hold, and the set of free states are not convex. However, convexity is for obvious reasons a convenient mathematical property to have and a natural property in some contexts. 
We therefore conclude with our last postulate (keeping in mind that there are QRTs that do not satisfy this assumption, and for which our main result cannot be applied):

\textbf{Postulate V}: {\it Each $\mathcal{F}_{\bf{m}}$ is a convex set.}

Every state that is not in $\mathcal{F}$ is considered a resource. Since $\mF$ is closed, the set of resource states is open. This means that resource states can be arbitrarily close to the set of free states and therefore motivate the notion
of highly resourceful states (those that are far from the set of free state) and weakly resourceful states (those that are very close to the set of free states). Indeed, this geometrical way to measure the resourcefulness of the states lead to a unique measure of resourcefulness in asymptotically reversible resource theories.

The set of free operations are the set of all possible operations given the restrictions at hand. The type of restrictions (and therefore the free operations) can vary drastically from one resource theory to another.
Hence, it is hard to imagine a general resource theory unifying all resource theories into a single framework.
Nevertheless, there is a general statement on the set of free operations that must hold true in all resource theories, and can be considered as the main characteristic of
a resource theory:

\textbf{The free operations postulate} (FOP): {\it The set of free operations can not generate a resource; they can not convert free states into resource states.}

Note that clearly free operations can convert one resource state into another. 
The intuition is that free operations can not convert a resource state into a more resourceful state.
However, the term ``more" resourceful implies a total order or hierarchy of resources.
Such a total order does not exists in general. In fact, in most cases it is a partial order that determines the hierarchy of resources. This kind of partial hierarchy varies a lot from one resource theory to another
and therefore can not be postulated in general terms. The only distinction that we can make here is between resource states and non-resource states. 

We denote by $NR$ the set of all completely positive maps that satisfies the FOP; i.e. $NR$ is the set of resource non-generating operations. We also denote by $NR^{\textbf{m}}$ the elements of $NR$ acting on $\mathcal{D}(\mathrm{H}_{{\bf m}})$.
Since the FOP is the only constraint on the elements of $NR$, the set $NR$ is bigger than or equal to the set of allowed operations. In fact, $NR$ is the maximal possible 
set of free operations in any non-trivial QRT.

Any measure or quantifier of the resource must be monotonically non-increasing under the action of free/allowed operations. This is a necessary condition if the
measure is to have operational significance (that is, it quantifies the
optimal figure of merit for some task that requires the resource for its implementation). 
If a measure is also monotonically non-increasing under any element of $NR$ 
then it is a resource measure for all QRTs with the same set $\mF$ of free states. 
Since the set of free states $\mathcal{F}_{\bf{m}}$ is convex and closed, it is well known that 
one can define a class of geometric resource quantifiers that are monotonic under $NR$ 
and that are based on the distance of the resource from the set of free states. 

The distance, in many resource theories, is measured by a \emph{contractive metric} (see e.g.~\cite{PV07,HHH09,VP,VP97,GMN14}) on the quantum states; that is, a metric $\mathcal{C}$ that assigns to two quantum states $\rho$ and $\sigma$, on the same underlying
Hilbert space, a non-negative real number $\mathcal{C}(\rho,\sigma)$ 
such that every completely positive, trace-preserving map $\Lambda$ is a contraction, i.e.
\be\label{cont}
   \mathcal{C}\left(\strut\Lambda(\rho),\Lambda(\sigma)\right)\leq \mathcal{C}(\rho,\sigma).
\ee
Then, any measure $M: \mathcal{D}(\mathrm{H}_{{\bf m}})\to\mathbb{R}_+$ 
$$
M(\rho):=\inf_{\sigma\in\mathcal{F}_{\bf{m}}}\mathcal{C}(\rho,\sigma)\;,
$$
where $\mathcal{C}$ is a contractive metric, is a resource quantifier and in particular non-increasing under any map in $NR^{\textbf{m}}$ (see e.g.~\cite{PV07,HHH09}). This can also be seen from that fact that
for any $\Lambda\in NR^{\textbf{m}}$ we have $\Lambda(\mathcal{F}_{\bf{m}})\subset\mathcal{F}_{\bf{m}}$ so that
\begin{align*}
M\left(\Lambda(\rho)\right) =\inf_{\sigma\in\mathcal{F}_{\bf{m}}}\mathcal{C}\left(\Lambda(\rho),\sigma\right)
&\leq \inf_{\sigma\in\mathcal{F}_{\bf{m}}}\mathcal{C}\left(\Lambda(\rho),\Lambda(\sigma)\right)\\
&\leq\inf_{\sigma\in\mathcal{F}_{\bf{m}}}\mathcal{C}\left(\rho,\sigma\right)=M(\rho)\;.
\end{align*}
where we have used~\eqref{cont} in the second inequality. 

The {\it relative entropy of a resource} is defined in a similar way as (see~\cite{VP,VP97} for the original definition in entanglement theory)
$$
E(\rho):=\inf_{\sigma\in\mathcal{F}_{\bf{m}}} S\left(\rho\|\sigma\right)\;.
$$
where $S(\rho\|\sigma)={\rm Tr}[\rho(\log\rho-\log\sigma)]$ is the relative entropy (which is not a metric).
Since $S(\rho\|\sigma)$ is contractive, the relative entropy of a resource is a monotone. This measures has many useful properties~\cite{VP,VP97,VPJK} and in particular is known to behave smoothly in the asymptotic regime when considering arbitrarily large number of copies of a quantum system~\cite{PV07,HHH09}. In this case, its variant,
the \emph{regularized} relative entropy of a resource is defined as
$$
E^\infty(\rho)\equiv\lim_{n\to\infty}\frac{1}{n}E(\rho^{\otimes n})\;,
$$
which play a key role in many quantum resource theories.

Lastly, we will be using in the definition below the robustness monotone.
The measure of robustness~\cite{VT99,AN03,BP08} in entanglement theory measures the amount of noise that can be added
to an entangled state before it becomes unentangled
(separable). This measure can be easily generalized to any resource theory as follows.
Let $\rho\in\mathcal{D}(\mathrm{H}_{{\bf m}})$.
Then, the (global) \emph{robustness} of $\rho$ 
is defined by
\be\label{pi}
\mathcal{R}(\rho):=\min_{\pi\in\mathcal{D}(\mathrm{H}_{{\bf m}})}\left\{s\geq 0\;:\;\frac{\rho+s\pi}{1+s}\in\fm\right\}
\ee
Both the robustness and the relative entropy measure are monotones under $NR$. They are also both convex and faithful (see e.g.~\cite{PV07,HHH09}) in the sense that they are zero iff $\rho\in\fm$. 

Since we focus here on resource manipulation in the limit of arbitrarily many copies of the state in question, we define an even larger class of maps than $NR$, those that are not  generating resources only in the asymptotic limit. 
For this purpose, we first define $\varepsilon$-resource non-generating operations.
\begin{definition}
Let $\Lambda:\dm\to\mathcal{D}(\mathrm{H}_{{\bf m}'})$ be a quantum operation. We say that $\Lambda$ is an $\varepsilon$-resource non-generating operation if for every free state $\sigma\in\dm$,
$
\mathcal{R}(\Lambda(\sigma))\leq\varepsilon\;.
$
We denote the set of $\varepsilon$-resource non-generating maps by $NR(\varepsilon)$.
\end{definition}
An asymptotically resource non-generating operation is then defined by a sequence of trace-preserving CP maps $\Lambda_n:\mathcal{D}(\mathrm{H}_{{\bf m}}^{\otimes n})\to\mathcal{D}(\mathrm{H}_{{\bf m'}}^{\otimes n})$, with $n\in\mbN$, such that $\Lambda_n$ is an $\varepsilon_n$-resource non-generating operation and $\lim_{n\to\infty}\varepsilon_n=0$. Finally, the optimal rate of converting (by asymptotically resource non-generating operations) $n$ copies of a resource state $\rho$ into $m$ copies of another resource $\sigma$  is defined by:
\begin{align}
&R(\rho \rightarrow \sigma) :=\min 
\nonumber \\
& \left \{ \frac{m}{n} \hspace{0.1 cm} : \hspace{0.1 cm}  \lim_{n \rightarrow \infty}  \left(\min_{\Lambda \in NR(\varepsilon_n)} \Vert \Lambda(\rho^{\otimes n}) - \sigma^{\otimes m}\Vert_1  \right) = 0 \right \},
\end{align}
with $\lim_{n \rightarrow \infty} \varepsilon_n = 0$.
With these definitions and notations we are ready to present the main result:
\begin{theorem}
Consider a QRT with a set of free states $\mathcal{F}$. If $\mF$ satisfies the 5 postulates above then
the regularized relative entropy of a resource can be expressed as 
\begin{align} \label{robustnessequaltoE}
&E^\infty(\rho) = \min_{ \{ \rho_n \in {\cal D}(\mathrm{H}_{{\bf m}}^{\otimes n}) \}  } \nonumber\\
&\left\{\lim_{n \rightarrow \infty}  \frac{ \log(1 + \mathcal{R}(\rho_n))}{n}   : \Vert \rho_n - \rho^{\otimes n} \Vert_1 \rightarrow 0\right\},
\end{align}
and for every $\sigma$ such that $E^\infty(\sigma) > 0$,
\begin{equation} \label{reversibility}
R(\rho \rightarrow \sigma) = \frac{E^\infty(\rho)}{E^\infty(\sigma)}.
\end{equation}
\end{theorem}
\begin{remark}
Eq.(\ref{reversibility}) in the theorem above identifies the regularized relative entropy as the `unique'
measure of a resource in the asymptotic limit. That is, there is a single function, $E^{\infty}$, that determines the rate of (reversible) conversion of many copies of $\rho$ to many copies of $\sigma$ under non-resource generating operations. Note however that
the proof of the theorem above can not follow directly from its analog in entanglement theory~\cite{BP08}.
Unlike entanglement theory, that have a unique ``golden" unit such as the Bell/Singlet state, general QRTs may have many such units, and more precisely, can have many inequivalent maximal resource states. For this reason, obtaining also general results in the single shot case, similar to the ones in~\cite{BND} for single shot entanglement theory, are far from being trivial and a subject for further study.
\end{remark}

The proof is partly based on a recent generalization of the quantum Stein's Lemma~\cite{BFP10},
which can be described in terms of the following property of QRTs.
\begin{definition}
Consider a QRT with a set of free states $\mathcal{F}$ and denote by $\mathcal{F}_n$ the set of all free states  
in $\mathcal{D}\left(\mathrm{H}_{{\bf m}}^{\otimes n}\right)$ (here \textbf{m} is a fixed dimension vector).
We say that the QRT satisfies the \emph{exponential distinguishability property} (EDP) if there is a non-identically-zero function $f : D({\cal H}) \rightarrow \mathbb{R}_+$ such that for every resource state $\rho$ and $\varepsilon > 0$,
\begin{equation}
\lim_{n \rightarrow \infty} - \frac{\log(\beta_n(\rho, \varepsilon))}{n} = f(\rho),
\end{equation}
with
\begin{equation} \label{dis1}
\beta_n(\rho, \varepsilon)\equiv \min_{0 \leq A_n \leq I} \left( \beta^{(2)}(A_n) : \beta^{(1)}(A_n) \leq \varepsilon   \right),
\end{equation}
where
$
\beta^{(2)}(A_n) \equiv \max_{\omega_n \in\mathcal{F}_n} \text{tr}(\omega_n A_n)
$ and
$\beta^{(1)}(A_n) \equiv\text{tr}(\rho^{\otimes n} (I - A_n))$.
\end{definition}

In~\cite{BFP10} it was shown that if the set $\mathcal{F}$ satisfies the 5 postulates then any such resource theory satisfies the EDP with $f = E^{\infty}$ being the regularized relative entropy of a resource. Furthermore, it was also shown in Proposition II.1 of~\cite{BFP10} that in this case the relative entropy of a resource can be expressed as in Eq.~\eqref{robustnessequaltoE}. 

To prove Eq. (\ref{reversibility}), consider the sequence of maps 
\begin{equation}
\Lambda_n(X) := \text{tr}(A_n X) \sigma_{n} + \text{tr}((I - A_n)X) \pi_n.
\end{equation}
In the equation above $\sigma_{n}$ is chosen such that both
$$
\Vert \sigma^{\otimes n \frac{E^\infty(\rho)}{E^\infty(\sigma)}} - \sigma_n \Vert_1 \rightarrow 0\;;\;
\lim_{n \rightarrow \infty} \frac{\log(1 + \mathcal{R}(\sigma_n))}{n} = E^\infty(\sigma)
$$
and $\pi_n$ is taken to be the optimal state in Eq.~\eqref{pi} for $\sigma_n$; that is, 
\begin{equation}
\frac{1}{1 + \mathcal{R}(\sigma_n)} \left(  \sigma_n + \mathcal{R}(\sigma_n) \pi_n  \right) \in {\cal F}_{ \left\lceil n \frac{E^\infty(\rho)}{E^\infty(\sigma)} \right\rceil}
\end{equation}
The sequence of POVMs $\{ A_n, I - A_n \}$ is chosen as the optimal one for $\rho$ in~(\ref{dis1}) with $\varepsilon_n \rightarrow 0$. With these choices we get
\begin{equation}
\Vert \Lambda_n(\rho^{\otimes n}) - \sigma^{\otimes n\frac{E^\infty(\rho)}{E^\infty(\sigma)}} \Vert_1 \rightarrow 0,
\end{equation}
so that indeed the rate is $E^\infty(\rho) / E^\infty(\sigma)$. It is left to show that 
$\{ \Lambda_n \}_n$ is asymptotically non-resource generating. Indeed,
since for every $\delta > 0$ and large enough $n$
\begin{equation}
\max_{\omega \in {\cal F}} \text{tr}(A_n \omega) \leq 2^{- n (E^\infty(\rho) - \delta)},
\end{equation}
and since
\begin{equation}
{\cal R}(\sigma_n) = 2^{n E^\infty(\rho)} - 1
\text{ and }
{\cal R}(\pi_n) \leq 1 / {\cal R}(\sigma_n)
\end{equation}
we find that indeed
$
\lim_{n \rightarrow \infty} \max_{\omega_n \in {\cal F}_n} {\cal R}(\Lambda_n(\omega_n)) = 0
$.

To summarize, we have shown that under 5 very reasonable assumptions on the set of free states $\mathcal{F}$,
a QRT is asymptotically reversible if the set of free operations is maximal. 
This does not mean that if the set of free operation is not maximal the theory is necessarily non-reversible. For example, the resource theory of pure bipartite entanglement is asymptotically reversible under LOCC which is a strictly smaller set than non-entangling operations. Yet, reversibility under LOCC no longer hold in the theory of mixed or multipartite entanglement. 
In such cases, where the set of free operations is not maximal, our results indicate 
how much one has to increase the set of allowed operations to achieve reversibility. 

In~\cite{HOO2002} it was shown that the regularized relative entropy of a resource is the unique asymptotic rate of any reversible QRT. This is consistent with our results, and also explains why the relative entropy of a resource plays a key role in many QRTs, such as the resource theory of entanglement, non-uniformity~\cite{GMN14}, athermality~\cite{BHO13,BHN13,HO13,FDO12,GMN14,NG14}, coherence (as defined in~\cite{BCP14}), stabilizer computation~\cite{VMG14}, and contextuality~\cite{GHH14}. The results presented here, however, can not be applied directly to \emph{all} QRTs since for example we considered only finite dimensional Hilbert spaces, and especially it is not applicable for the resource theory of Non-Gaussianity (also the set of Gaussian states is not convex~\cite{BL05}).

While the restriction to finite dimensions is a significant one we
expect that under a suitable energy constraint, our main result can be extended
to infinite dimensional systems. However
since it would require a long technical argument to establish it, we are
leaving it to future work. 
Finally, our main theorem also cannot be applied, in a straightforward manner, to the resource theory of asymmetry  since in that theory the regularized relative entropy of asymmetry is zero~\cite{GMS09}. We believe that this can be resolved by proper rescaling of the relative entropy of asymmetry (as was shown for a special case of $U(1)$-symmetry in~\cite{SG12}) and is also left for future work. 

\emph{Acknowledgments:---}
F.G.S.L.B. is supported by an EPSRC Early Career Fellowship. G.G. research is supported by NSERC. We thank Michal Horodecki, Jonathan Oppenheim, Martin Plenio and Rob Spekkens for interesting and stimulating discussions.

\end{document}